\documentclass[aps,prd,twocolumn,groupedaddress]{revtex4}
\usepackage{graphicx}
\usepackage{dcolumn}
\usepackage{bm}
\usepackage{amsmath}
\usepackage{amssymb}
\usepackage{color}
\usepackage{slashed}

\begin{document}

\title{How can every physicist become a Bohmian}

\author{Hrvoje Nikoli\'c}\email{hnikolic@irb.hr}
\affiliation{Theoretical Physics Division, Rudjer Bo\v{s}kovi\'{c} Institute, \\
Bijeni\v{c}ka c.~54, HR-10002 Zagreb, Croatia}

\begin{abstract}
The Bohmian interpretation of quantum mechanics is controversial because it attributes deterministic trajectories to microscopic particles. We propose a reformulation that sidesteps this controversy: rather than applying Bohmian mechanics at the microscopic level, we recast it as an effective deterministic description of the macroscopic pointer of a measuring apparatus, with no claims about the fundamental nature of the microscopic world. In this form, the Bohmian trajectory becomes a practical tool for visualizing the emergence of a definite measurement outcome, broadly acceptable to physicists who would otherwise reject the Bohmian program.
\end{abstract}


\maketitle

\section{Introduction} 
\label{SEC1}

The Bohmian interpretation
\cite{bohm,book-bohm,book-hol,book-durr,oriols,tumulkaAJP,nikBBB}
of quantum mechanics (QM), in its usual formulation, claims that every microscopic particle, like an electron, has a deterministic trajectory guided by the pilot wave. In such a form the Bohmian interpretation is controversial, most physicists do not find it very convincing (see e.g. \cite{passon,nikBBB} and references therein). In addition, even though the Bohmian interpretation makes the same measurable predictions
as standard QM \cite{bohm,book-bohm,book-hol,book-durr,oriols,tumulkaAJP,nik_IBM},
most physicists do not really understand how exactly the Bohmian interpretation does that.

To fix these deficiencies, in this paper we entirely reformulate Bohmian mechanics (BM), by reducing it to the bare minimum that every physicist can easily understand and find much more plausible. In this reformulation we do not talk about microscopic particle trajectories. In fact, we say nothing about the microscopic world except the bare minimum that most physicists already know from any basic textbook on QM. Instead, we shift the focus to something that in QM textbooks is entirely neglected, we talk about the behavior of the macroscopic apparatus.

But to make the analysis simple, we do not talk about the details of realistic apparatuses. The details would only hide the true essence of the apparatus, while its essence is really simple. Instead, to see how simple it is, we strip out all the inessential details and use a very simple model of the measuring apparatus; we use the von Neumann model \cite{neumann,steinberg}. The von Neumann model has been studied in the Bohmian context before \cite{bohm,book-bohm,book-hol}, but our analysis is different because we study a model in which the pointer of the apparatus is the {\it only} thing that has a Bohmian trajectory.

The paper is organized as follows. The basic principles of our theory are presented in Sec.~\ref{SEC2}, while details of the measuring process are explained in Sec.~\ref{SEC3}. The nonlocal aspects of the theory are discussed in Sec.~\ref{SEC4}, and finally, the conclusions are drawn in Sec.~\ref{SEC5}.
In the Appendix, some formal expressions that play an important role in the paper are represented in a more abstract form.

\section{The theory in a nutshell}
\label{SEC2}

\subsection{The von Neumann model in standard QM}

In the von Neumann model \cite{neumann,steinberg}, the process of measurement is described by the Hamiltonian
\begin{equation}\label{H}
H=g(t)Ap,
\end{equation}
where $A$ is the measured observable, $p$ is the momentum of the macroscopic pointer of the measuring apparatus, and $g(t)$ is a time-dependent coupling function. Later we shall explain in more detail why the Hamiltonian has this form, but for now let us just take this Hamiltonian at face value. The corresponding Schr\"odinger equation is
\begin{equation}
H|\Psi(t)\rangle=i\hbar\partial_t|\Psi(t)\rangle .
\end{equation}

Let $\{ |k\rangle \}$ be some orthonormal basis, $\langle k|k'\rangle=\delta_{kk'}$, for the Hilbert space in which the measured observable $A$ lives, and let $\{ |x\rangle \}$ be the position basis for the macroscopic pointer. Thus the basis for the full Hilbert space are the states $\{ |k,x\rangle \}$, where $|k,x\rangle \equiv |k\rangle|x\rangle$. With this basis we can define the wave function
\begin{equation}
\Psi(k,x,t)=\langle k,x|\Psi(t)\rangle.
\end{equation}
But we want to introduce a more compact notation that hides the dependence on $k$, so we define the multi-component wave function $\Psi(x,t)$ with components  $\Psi(k,x,t)$. Assuming that $k$ takes values from a discrete set $k\in\{k_1,k_2,\ldots\}$, one can think of $\Psi(x,t)$ as a column
\begin{equation}\label{column}
\Psi(x,t)=\left(
\begin{matrix}
\Psi(k_1,x,t) \\
\Psi(k_2,x,t) \\
\vdots
\end{matrix}
\right).
\end{equation}
In this picture, the operator $A$ is a matrix with components $A(k',k)=\langle k'|A|k\rangle$. A familiar example is the case of the spin operator in the $z$-direction, $A=\frac{\hbar}{2}\sigma_z$, with $\sigma_z$ being the $2\times 2$ Pauli matrix, in which case (\ref{column}) is a 2-component spinor. But it is important to keep in mind that $x$ is not the position of the particle with spin. Instead, it is the position of the macroscopic pointer.

Now, representing the momentum operator with $p=-i\hbar\partial_x$, the Schr\"odinger equation with the Hamiltonian (\ref{H}) can be written as
\begin{equation}\label{1}
-gA\partial_x\Psi(x,t)=\partial_t\Psi(x,t).
\end{equation}
Since $A$ is an observable, it must be a hermitian operator $A^{\dagger}=A$. The hermitian conjugate of (\ref{1}) is thus
\begin{equation}
-g \partial_x\Psi^{\dagger}(x,t)A=\partial_t\Psi^{\dagger}(x,t),
\end{equation}
so
\begin{equation}
\partial_t(\Psi^{\dagger}\Psi)=-g\partial_x(\Psi^{\dagger}A\Psi) .
\end{equation}
This can be written as a continuity equation
\begin{equation}\label{cont}
\partial_t\rho + \partial_x(\rho v)=0,
\end{equation}
where
\begin{equation}
\rho(x,t)\equiv\Psi^{\dagger}(x,t)\Psi(x,t),
\end{equation}
\begin{equation}\label{v}
v(x,t)\equiv\frac{g(t)\Psi^{\dagger}(x,t)A\Psi(x,t)}{\Psi^{\dagger}(x,t)\Psi(x,t)}.
\end{equation}
The $\rho(x,t)$ is the probability density for the pointer to have the position $x$ at time $t$, so the continuity equation provides the conservation of probability. In particular, if $\Psi(x,t)$ is an eigenstate of $A$
\begin{equation}
A\Psi(x,t)=a\Psi(x,t),
\end{equation}
with $a$ being a real number, then (\ref{v}) reduces to
\begin{equation}\label{va}
v(t)=g(t)a.
\end{equation}

\subsection{The Bohmian interpretation}

So far, the analysis above has been based on standard QM, there was nothing specifically Bohmian about it. Now it is time to introduce the Bohmian interpretation of the pointer with the position $x$. Even though the pointer was treated quantum mechanically, we recall that it is a macroscopic object. Thus, intuitively, one expects that somehow it should have a classical-like trajectory $X(t)$. But how can that be compatible with QM?

The clue is the continuity equation (\ref{cont}). Even though it was derived from QM, it also has a classical-like interpretation. In such an interpretation, the quantity $v(x,t)$ is the actual velocity of a particle at position $x$, namely
\begin{equation}\label{dotx}
\dot{X}(t)=v(X(t),t).
\end{equation}
The Bohmian interpretation is nothing but taking (\ref{dotx}) seriously; the Bohmian interpretation postulates that the pointer has the trajectory $X(t)$ with the velocity obeying (\ref{dotx}). This trajectory is deterministic, as one expects from a macroscopic object. And yet, one can describe it probabilistically, in a way one describes probabilities in classical physics. Namely, if one does not know the exact initial condition $X(0)$, then this lack of knowledge leads to an effective probabilistic description. More specifically, if we study a statistical ensemble of many possible initial positions $X(0)$, and if the probability distribution of the initial positions is given by the quantum-mechanical distribution $\rho(X(0),0)$, then the continuity equation provides that the probability distribution is given by $\rho(X(t),t)$ at any time $t$. In this way, the macroscopic pointer has a deterministic trajectory, and yet, at each time $t$, the probability that it has the position $X$ is described by  QM.

Note that, in general, the Bohmian classical-like trajectory $X(t)$ is not identical to the truly classical trajectory $X_{\rm class}(t)$. The latter is obtained when the Hamiltonian (\ref{H}) is interpreted classically, $H_{\rm class}=gA_{\rm class}p_{\rm class}$, leading to the classical Hamilton equation of motion
\begin{equation}\label{class}
\dot{X}_{\rm class}=\frac{\partial H_{\rm class}}{\partial p_{\rm class}}=gA_{\rm class} .
\end{equation}
However, when the quantum state is an eigenstate of $A$, then (\ref{va}) and (\ref{dotx}) give
\begin{equation}
\dot{X}=ga ,
\end{equation}
which is identical to the classical equation (\ref{class}) with $A_{\rm class}=a$.

\section{Details of the measuring process}
\label{SEC3}

\subsection{Quantum mechanics without the measuring apparatus}

Quantum theory is usually presented without any explicit analysis of the measuring apparatus. The probability that an observable will have a certain value is computed by the Born rule, which does not make any explicit reference to the apparatus. To motivate the rest of the analysis in this paper, in the present subsection we present some aspects of quantum theory in this usual form, without the apparatus, and explain why such a formulation is not completely satisfying.

As we said, the measured observable $A$ lives in the Hilbert space with the basis $\{|k\rangle\}$. To simplify the analysis, we choose the basis which consists of eigenstates of $A$,
\begin{equation}\label{eigen}
A|k\rangle=a_k|k\rangle,
\end{equation}
with $a_k$ being the eigenvalue, so $A$ can be written in the spectral form
\begin{equation}\label{spectral}
A=\sum_k a_k |k\rangle\langle k| .
\end{equation}
To make the analysis rather general, we study an observable that can have a degenerate spectrum. This means that some of the eigenvalues may be the same, namely, for some $k\neq k'$, $a_k=a_{k'}$. Hence we introduce the notation $k=(d,l)$ where $d$ labels different eigenvalues, while $l$ labels different eigenstates with the same eigenvalue labeled by $d$. Thus (\ref{eigen}) can be rewritten as
\begin{equation}\label{eigen2}
A|d,l\rangle=a_d|d,l\rangle,
\end{equation}
where all $a_d$ are different, and (\ref{spectral}) can be rewritten as
\begin{equation}\label{spectral2}
A=\sum_d a_d \pi_d ,
\end{equation}
where $\pi_d$ are projectors
\begin{equation}
\pi_d = \sum_l |d,l\rangle\langle d,l| ,
\end{equation}
obeying $\pi_d\pi_{d'}=\pi_d\delta_{dd'}$ and $\sum_d \pi_d=\sum_k |k\rangle\langle k| =1$.

A general state in this Hilbert space is a superposition
\begin{equation}\label{psi}
|\psi\rangle = \sum_k c_k |k\rangle = \sum_{d,l} c_{dl} |d,l\rangle ,
\end{equation}
where the coefficients $c_k$ obey the normalization condition
\begin{equation}
\sum_k |c_k|^2=\sum_{d,l} |c_{dl}|^2=1.
\end{equation}

Also note that the state $|\psi\rangle$ can be viewed as a vector represented by a column, as in (\ref{column}), so (\ref{psi}) can also be written as
\begin{equation}\label{psi2}
\psi = \sum_k c_k \psi_k = \sum_{d,l} c_{dl} \psi_{dl} ,
\end{equation}
where $\psi_k=\psi_{dl}$ are the basis column vectors. In this notation, the normalization $\langle\psi|\psi\rangle=1$ is written as $\psi^{\dagger}\psi=1$, and similarly for the basis vectors, $\psi^{\dagger}_k\psi_k=\psi^{\dagger}_{dl}\psi_{dl}=1$.

Now the Born rule can be stated concisely as follows: The probability that the observable $A$ will have the value $a_d$ is given by the formula
\begin{equation}\label{born}
p_d=\sum_{l} |c_{dl}|^2 .
\end{equation}

The Born rule is defined completely in terms of objects in the Hilbert space of the observable, without any reference to a measuring apparatus. This creates an impression that the probability is an intrinsic property of the system itself, irrespective of measurement. However, this cannot be quite true. If we consider probabilities of two mutually incompatible observables, for example position and momentum, or spin in two different directions, the Born rule does not give a joint probability for both observables. One must consider the probability of either one observable or the other, not of both observables simultaneously. The standard interpretation of this fact is that the probability somehow depends on the measurement itself. Does it mean that, in a full analysis, one must also take into account a measuring apparatus? For practical purposes one does not, because the Born rule as described above is enough to compute all the probabilities that can be measured, and these probabilities turn out to agree with observations. But for conceptual understanding this is not fully satisfactory. To understand how the probabilities depend on the measurement, one wants to understand in more detail how the measuring apparatus affects probabilities. After all, the measuring apparatus is a physical object too, so physics should be able to describe it. That is what we study in the rest of this section.

\subsection{Basic ingredients of a measuring apparatus}

In Sec.~\ref{SEC2} we studied the von Neumann Hamiltonian (\ref{H}) without explaining where does it come from and how it describes a measurement. Now we explain these things in more detail.

Suppose that you want to measure some observable $A$. It can by anything. For example, it can be a position of a particle, or its momentum, or energy, or spin, whatever you like. Or it can be some collective property of a many-particle system. Or it can be some quantity in field theory, perhaps some function of electric and magnetic field. It does not even need to be a quantum observable, it can be a purely classical quantity. All we need to assume is that $A$ is some quantity that can be measured. The beauty of the von Neumann model lies in the fact that it does not care what the observable $A$ really is, the theory works the same way for any $A$.

To measure $A$, we need some macroscopic apparatus. The essential part of any apparatus is the thing which shows us the final result of measurement. In a realistic digital apparatus it can be some display with digits. But in our simple model of the apparatus this thing will be a needle with the position $x$. In the literature, such a position of a needle is usually called a pointer. This pointer with position $x$ is the main subject of our analysis, all we want to understand is what happens with $x$.

In addition to the measured observable $A$ and the pointer $x$, we need some interaction between them. For simplicity we assume that the interaction lasts a finite time $\tau$, so we describe the interaction with a time-dependent coupling function $g(t)$, which is nonzero during a finite time from $t=0$ to $t=\tau$. With these three ingredients, $A$, $x$ and $g(t)$, von Neumann proposed that the coupling is described by the interaction Hamiltonian
\begin{equation}\label{H3}
H=g(t)Ap ,
\end{equation}
where $p$ is the momentum of the macroscopic pointer, canonically conjugated to its position $x$.

To further simplify the analysis, it is assumed that during the measurement the interaction part of the Hamiltonian dominates, so that all other contributions to the Hamiltonian can be neglected. Thus the interaction Hamiltonian $H$ above is treated as it was the full Hamiltonian. In this way it is easy to see how that Hamiltonian describes the measurement.

\subsection{Classical measurement}

Let us first see how it works classically. We have already discussed it briefly in Sec.~\ref{SEC2}, but here we shall present more details and we will use a slightly different notation. In particular, the classical trajectory will now be denoted by $x(t)$, which previously was denoted by $X_{\rm class}(t)$. The Hamilton equation of motion for the pointer is
\begin{equation}\label{3.1}
\dot{x}=\frac{\partial H}{\partial p}=gA .
\end{equation}
Hence
\begin{equation}
x(t)=\int_0^t dt \, g(t)A,
\end{equation}
so after the interaction, the final position of the pointer is
\begin{equation}\label{3.2}
x=GA,
\end{equation}
where $G\equiv \int_0^{\tau} dt \, g(t)$ and it is assumed that $A$ does not change during the measurement. This assumption can be justified in two ways. First, $A$ is clearly conserved by the evolution governed by the Hamiltonian (\ref{H3}), because this Hamiltonian itself is proportional to $A$, so the Poisson bracket $\{H,A\}$ vanishes. Since we assume that other contributions to the Hamiltonian can be neglected during the measurement, any change of $A$ due to other contributions can be neglected. Second, even if other contributions to the Hamiltonian are not negligible, we can assume that $\tau$ is a short time so that $A$ does not have time to change much during the measurement. But these assumptions are in no way essential, their only purpose is to simplify the model of measurement as much as possible. The upshot of this simple model is to establish the correlation between the measured quantity $A$ and the pointer position $x$, described by (\ref{3.2}).

\subsection{Quantum measurement in the Heisenberg and Schr\"odinger pictures}

In the Heisenberg picture, the quantum equation of motion is analogous to the classical equation of motion. The basic assumption is that QM is a universal theory, valid not only for the microscopic world, but also for the macroscopic pointer. Hence $p$ in the Hamiltonian above is viewed as a quantum operator, obeying the commutation relation $[x,p]=i\hbar$. Therefore, in the Heisenberg picture, the quantum version of (\ref{3.1}) is the operator equation
\begin{equation}\label{3.3}
\dot{x}=-\frac{i}{\hbar}[x,H]=gA .
\end{equation}

In the Schr\"odinger picture, the wave function evolves according to
\begin{equation}
\Psi(x,t)=e^{-\frac{i}{\hbar}\int_0^t dt\, H(t)} \Psi(x,0) ,
\end{equation}
so after the interaction the final state is
\begin{equation}\label{psitau}
\Psi(x,\tau)=e^{-\frac{i}{\hbar}GAp} \Psi(x,0) .
\end{equation}

To continue, we have to specify the initial state $\Psi(x,0)$. That is what we study in the following subsections. For convenience, in the rest of the analysis we work in the Schr\"odinger picture.

\subsection{The case of a well defined value of the observable}
\label{SECl}

Suppose that, in the initial state, the observable $A$ has a well defined value $a_d$. This means that the corresponding state has the form
\begin{equation}\label{psid}
\psi_d=\sum_l c_{dl}\psi_{dl} .
\end{equation}
The full initial state state, including the pointer of the apparatus, is assumed to be a product state
\begin{equation}\label{init}
\Psi(x,0)=\psi_d\phi(x) ,
\end{equation}
where $\phi(x)$ is the initial wave function of the pointer. We assume that $\phi(x)$ is a narrow wave packet (e.g. a Gaussian), which has a peak at $x=0$ and is negligible out of the small interval $\sigma_0$ around $x=0$. Hence the normalization  formula $\int_{-\infty}^{\infty} dx\, |\phi(x)|^2=1$ can be approximated with
\begin{equation}\label{intsigma0}
\int_{\sigma_0} dx\, |\phi(x)|^2 \simeq 1.
\end{equation}
The initial wave function (\ref{init}) can thus be written as
\begin{equation}
\Psi(x,0)=\sum_l c_{dl} \Psi_{dl}(x) ,
\end{equation}
where $\Psi_{dl}(x)\equiv\psi_{dl}\phi(x)$ and $A\Psi_{dl}(x)=a_d\Psi_{dl}(x)$. Hence the final state (\ref{psitau}) after the measurement is
\begin{equation}\label{psitau0}
\Psi(x,\tau)=\sum_l c_{dl} e^{-\frac{i}{\hbar}Ga_dp} \Psi_{dl}(x) .
\end{equation}
Since $p=-i\hbar\partial_x$, we have
\begin{eqnarray}
e^{-\frac{i}{\hbar}Ga_{d}p} \Psi_{dl}(x) & = & e^{-Ga_{d}\partial_x} \Psi_{dl}(x)
\nonumber \\
& = & \sum_{n=0}^{\infty} \frac{(-Ga_{d})^n}{n!}\partial^n_x \Psi_{dl}(x)
\nonumber \\
& = & \Psi_{dl}(x-Ga_d)
\nonumber \\
& = & \psi_{dl}\phi(x-Ga_d) ,
\end{eqnarray}
where in the transition from the second to the third line we have recognized that the second line is a Taylor expansion of the third line. (This can be seen more easily by starting from Taylor expansion in a more familiar form $f(y)=\sum_n [(y-x)^n/n!]\partial_x^n f(x)$ and then taking $y=x-Ga_d$.) In this way, (\ref{psitau0}) finally gives
\begin{equation}\label{psitau2}
\Psi(x,\tau)=\sum_l c_{dl} \psi_{dl}\phi(x-Ga_d) = \psi_d \phi(x-Ga_d) .
\end{equation}
Thus we see that, during the measurement, the initial wave function $\phi(x)$ of the pointer is translated in the $x$-space by $Ga_d$, having the same narrow shape as initially, but with the peak at $x=Ga_d$. This is fully analogous to the classical measurement, where the pointer shifts from the initial position $x=0$ to the final position $x=GA$, as obtained in (\ref{3.2}).

\subsection{The case of a superposition}
\label{SECdl}

The true distinction between classical and quantum measurements shows up when the initial state does not have a well defined value of the measured observable. In this case (\ref{psid}) generalizes to
\begin{equation}\label{psids}
\psi=\sum_{d,l} c_{dl}\psi_{dl} ,
\end{equation}
involving a superposition over the eigenstates with different values of $a_d$. Now (\ref{init}) generalizes to $\Psi(x,0)=\psi\phi(x)$, so the final state (\ref{psitau2}) generalizes to
\begin{equation}\label{psitau3}
\Psi(x,\tau)=\sum_{d,l} c_{dl} \psi_{dl}\phi(x-Ga_d) .
\end{equation}
Now the wave function of the macroscopic pointer itself is in the superposition of wave packets localized around different positions $x=Ga_d$. On the other hand, we observe a macroscopic pointer at only one position. How to explain that?

Standard QM postulates that the superposition (\ref{psitau3}) somehow ``collapses" to one of the terms with a definite $d$, namely, to one of the states of the form (\ref{psitau2}). The Bohmian interpretation, however, offers a different interpretation, much more classical in spirit. Each wave packet $\phi(x-Ga_d)\equiv\phi_d(x)$ is negligible out of the small interval $\sigma_d$ around $x=Ga_d$. We assume that $G$ is sufficiently large so that there is no overlap between the intervals $\sigma_d$ with different $d$, so the corresponding wave functions obey
\begin{equation}
\phi_d(x)\phi_{d'}(x) \simeq 0 \;\;\; {\rm for} \;\;\; d\neq d' ,
\end{equation}
and (\ref{intsigma0}) implies
\begin{equation}\label{intsigmad}
\int_{\sigma_d} dx\, |\phi_d(x)|^2 \simeq 1.
\end{equation}
The Bohmian interpretation posits that the pointer has an actual position $X(t)$, so it can be in only one of the regions $\sigma_d$. So if the pointer ends up in the region $\sigma_d$, this effectively selects the corresponding term in (\ref{psitau3}) with definite $d$. This has the same effect on the pointer as if the wave function collapsed to this term, even though there was no actual collapse. In this way, the Bohmian interpretation explains the illusion of collapse.

What is the probability that the pointer will end up in the region $\sigma_d$? For this purpose we need to compute the probability density $\rho(x,\tau)=\Psi^{\dagger}(x,\tau)\Psi(x,\tau)$. Since $\psi^{\dagger}_{dl}\psi_{d'l'}=\delta_{dd'}\delta_{ll'}$, (\ref{psitau3}) gives
\begin{equation}\label{psitau4}
\rho(x,\tau)=\sum_{d,l} |c_{dl}|^2 |\phi_d(x)|^2 .
\end{equation}
Therefore the probability that the pointer will end up the region $\sigma_d$ is
\begin{equation}
p_d=\int_{\sigma_d} dx\,\rho(x,\tau) \simeq \sum_{l} |c_{dl}|^2 ,
\end{equation}
where we have used (\ref{intsigmad}). But this is the same result as the Born rule (\ref{born}). Thus the analysis above explains the Born rule in terms of positions of the macroscopic pointer.

\subsection{Should we take it seriously?}

The analysis above shows that the macroscopic Bohmian framework is mathematically consistent and reproduces the Born rule without postulating wave function collapse. But a natural question arises: should one regard the pointer trajectory $X(t)$ as a real physical trajectory, or merely as a convenient calculational fiction?

A word of caution is warranted here. The von Neumann model, while elegant and illuminating, is not a fully realistic description of an actual measuring apparatus. Real laboratory instruments contain enormously many degrees of freedom, electrons, nuclei, phonons, electromagnetic fields, all interacting in complicated ways. The Hamiltonian (\ref{H3}) captures only the essential coupling between the measured observable $A$ and a single collective degree of freedom, the pointer position $x$. It ignores the internal dynamics of the apparatus, thermal fluctuations, decoherence, and the coupling of the pointer to its environment. In a realistic setting, all of these play a role. The clean separation between the pointer and all other degrees of freedom, which made the mathematics in Secs.~\ref{SECl} and \ref{SECdl} so transparent, is an idealization.

Furthermore, the pointer is not truly a point particle. It is a macroscopic rigid body, or rather, it is an effective collective variable representing the position of a macroscopic needle, digit display, or some other readout device. Its quantum mechanical description in terms of a narrow wave packet $\phi(x)$ moving in one dimension is again an idealization. In reality, decoherence due to interaction with environmental degrees of freedom renders the pointer effectively classical on timescales far shorter than any relevant measurement time, so the quantum coherence of the pointer wave function is not an observable feature of the experiment in any direct sense.

None of this, however, undermines the purpose of the macroscopic Bohmian framework presented here. The trajectory $X(t)$ is best understood as an effective description, a tool for visualizing, in an intuitive and classically meaningful way, how a definite measurement outcome emerges from unitary quantum evolution. Within the von Neumann model, it does exactly that job: it shows, without invoking collapse, how the pointer ends up at one of the positions $Ga_d$, and how the probabilities of these outcomes reproduce the Born rule. This is precisely the conceptual gap that the standard formulation of QM, with its silent reliance on the Born rule applied to abstract Hilbert space vectors, leaves unfilled.

One should therefore think of the Bohmian pointer trajectory not as a claim about the ultimate microscopic ontology of the world, but as an effective macroscopic description of the pointer, valid for the purposes of building intuition about the measurement process. Much as the Navier–Stokes equations offer an effective description of a fluid without requiring one to track every molecule, the Bohmian guidance equation (\ref{dotx}) offers an effective description of the pointer without requiring one to commit to a specific view of microscopic particle trajectories. The theory is used here not as a fundamental ontology but as a computational and conceptual framework at the macroscopic level.

In this spirit, the macroscopic Bohmian framework can be broadly acceptable even to physicists who would otherwise be skeptical of the full Bohmian program. One need not believe that every electron in the universe has a hidden deterministic trajectory guided by a pilot wave. One need only accept the following modest set of claims: (i) the von Neumann Hamiltonian captures the essential physics of the coupling between a measured system and a macroscopic pointer; (ii) the pointer, being macroscopic, should admit a classical-like description; and (iii) the continuity equation (\ref{cont}), which follows directly from standard QM, provides just such a description through the guidance equation (\ref{dotx}). These are claims that virtually every physicist can accept without discomfort.

To summarize: the Bohmian pointer trajectory $X(t)$ should be taken seriously as an effective macroscopic description, not as a fundamental microscopic truth. It is a useful, internally consistent, and physically motivated way of visualizing how definite measurement outcomes arise from quantum mechanics, and it reproduces the correct quantum probabilities. Whether one chooses to extend this picture all the way down to the microscopic level, as the full Bohmian program does, is a separate question, and one that this paper deliberately leaves open.

\section{Nonlocality}
\label{SEC4}

So far we have studied a single measured system, described by the observable $A$, interacting with a single pointer $x$. In this setting, the Bohmian pointer trajectory offers a straightforward and self-contained picture. But quantum mechanics is well known to exhibit a richer and more subtle structure when two spatially separated systems are considered together: the phenomenon of entanglement. It is natural to ask how the macroscopic Bohmian framework extends to this case, and what it implies for the relationship between the two pointers. As we shall see, the answer brings us directly to nonlocality, as an unavoidable mathematical consequence of the guidance equation.

Suppose therefore that there are two spatially separated systems. In one of them the observable $A_1$ is measured by the pointer $x_1$, while in the other the observable $A_2$ is measured by the pointer $x_2$. All the results in this paper for one system are readily generalized to the case of two systems, so we sketch the main results without detailed derivations.

The Hamiltonian for two spatially separated systems is
\begin{equation}
H=H_1+H_2,
\end{equation}
where
\begin{equation}
H_i=g_i(t)A_ip_i ,
\end{equation}
for $i=1,2$. A general state without the apparatus has an entangled form
\begin{equation}
|\psi\rangle=\sum_{k_1}\sum_{k_2}c_{k_1k_2} |k_1\rangle|k_2\rangle \equiv \sum_k c_k |k\rangle ,
\end{equation}
where we have introduced a collective label
\begin{equation}
k=(k_1,k_2).
\end{equation}
Likewise, we introduce a collective notation for the pointer positions
\begin{equation}
x=(x_1,x_2),
\end{equation}
so that all equations can be written in the same form as for one system. The full wave function now has the form $\Psi(x,t)=\Psi(x_1,x_2,t)$, and the probability density $\rho=\Psi^{\dagger}\Psi$ obeys the continuity equation
\begin{equation}
\partial_t\rho +\sum_{i=1,2} \partial_{x_i}(\rho v_i)=0 ,
\end{equation}
where
\begin{equation}
v_i(x,t)\equiv\frac{g_i(t)\Psi^{\dagger}(x,t)A_i\Psi(x,t)}{\Psi^{\dagger}(x,t)\Psi(x,t)}.
\end{equation}
Thus the Bohmian equations of motion for the two pointers are $\dot{X}_i(t)=v_i(X(t),t)$, or more explicitly
\begin{equation}
\dot{X}_i(t)=v_i(X_1(t),X_2(t),t) .
\end{equation}
In this way we see that, in general, the velocity of any pointer at time $t$ depends on positions of all pointers at the same time $t$. This feature is nonlocality, typical for Bohmian mechanics.

This explicit nonlocality aligns directly with Bell’s theorem \cite{bell,bell_scholar,tumulka_assump},
which establishes that any completion of QM reproducing standard quantum predictions while maintaining objective, single outcomes must be nonlocal. In this framework, the underlying assumption is simply that macroscopic pointers of the laboratory measurement apparatuses always register definite positions, regardless of whether a conscious observation is taking place.

While full microscopic realism is often a point of contention in quantum foundations, framing the requirement purely at the macroscopic level reflects everyday laboratory reality: experiments ultimately yield unambiguous, single-valued pointer readings. Even for physicists who prefer an operational or instrumentalist viewpoint, describing the measurement process through definite macroscopic configurations serves as a consistent and logically sound model. Crucially, this model demonstrates that any framework providing a clear, objective account of how definite measurement outcomes emerge from quantum dynamics must be explicitly nonlocal, as mandated by Bell's theorem.

\section{Conclusion}
\label{SEC5}

The central message of this paper is simple: one does not need to be a committed Bohmian to find Bohmian mechanics useful. By shifting the focus from microscopic particle trajectories to the macroscopic pointer of a measuring apparatus, we have constructed a reformulation of Bohmian mechanics that is both mathematically transparent and conceptually undemanding.

The starting point is entirely standard. The von Neumann Hamiltonian (\ref{H}) is a well-established model of the measurement process, the Schrödinger equation is unmodified, and the continuity equation (\ref{cont}) for the pointer probability density follows as a straightforward consequence. The only additional step, and it is a modest one, is to interpret the velocity field $v(x,t)$ in that continuity equation as the actual velocity of the macroscopic pointer, giving the guidance equation (\ref{dotx}). This single interpretational move is enough to replace the mysterious collapse postulate with a concrete, deterministic trajectory, while recovering the Born rule as a natural consequence of the pointer ending up in one of the non-overlapping wave packets that form after the measurement interaction.

The resulting picture is not presented as a final or fundamental theory of nature. The von Neumann model is a deliberate simplification, and the pointer trajectory is best regarded as an effective description at the macroscopic level, one that provides physical intuition about how a definite outcome crystallizes out of a quantum superposition, without making strong commitments about the deeper structure of the microscopic world. Its value is conceptual clarity, not ontological completeness.

The analysis also reveals something that holds well beyond the specific framework adopted here. When the requirement of definite macroscopic outcomes is taken seriously, which is a mild requirement, since every real experiment ends with a pointer at a specific position, then the resulting dynamics is explicitly nonlocal, in full agreement with Bell’s theorem. This nonlocality is not an artifact of the Bohmian approach; it is a necessary feature of any theory that reproduces the quantum predictions while insisting that measurement outcomes are real and unambiguous.

We hope that the reformulation presented here lowers the barrier for physicists who have found the standard Bohmian program unappealing. The microscopic ontology of Bohmian mechanics, the claim that every particle always has a definite position guided by the universal wave function, may remain a matter of taste and philosophical conviction. But the macroscopic version developed here asks for nothing beyond what is already implicit in the everyday practice of experimental physics: that apparatuses have definite readings, and that these readings deserve a physical explanation.

\section*{Author Declarations}

The authors have no conflicts to disclose.
   
\section*{Acknowledgments}

This research was supported by the Croatian Science Foundation Project No. IP-2025-02-8625, {\it Quantum aspects of gravity}.

\appendix

\section{A more abstract formulation}
\label{SEcapp}

In this paper we represented the quantum state with a multi-component wave function (\ref{column}),
so the corresponding probability density was expressed as
$\rho(x,t)=\Psi^{\dagger}(x,t) \Psi(x,t)$. This can also be written as
\begin{eqnarray}
\Psi^{\dagger}(x,t) \Psi(x,t) &=& \sum_k \Psi^*(k,x,t) \Psi(k,x,t)
\nonumber \\
&=&  \sum_k \langle\Psi(t)|k,x\rangle\langle k,x|\Psi(t)\rangle
\nonumber \\
&=&  \sum_k \langle k,x|\Psi(t)\rangle \langle\Psi(t)|k,x\rangle
\nonumber \\
&=&  \sum_k \langle k,x|\rho(t)|k,x\rangle
\nonumber \\
&=& \langle x|{\rm Tr}_{\rm no\;pointer}\rho(t)|x\rangle ,
\end{eqnarray}
where $\rho(t)=|\Psi(t)\rangle \langle\Psi(t)|$ is the density matrix and ${\rm Tr}_{\rm no\;pointer}$
denotes the partial trace in the Hilbert space of all degrees of freedom that do not represent the pointer.
Likewise, the expression in the numerator of (\ref{v}) can be written as
\begin{equation}
\Psi^{\dagger}(x,t)A\Psi(x,t) =
 \langle x|{\rm Tr}_{\rm no\;pointer}A\rho(t)|x\rangle .
\end{equation}
Thus, instead of representing the state with the wave function, in such a more abstract formulation
the state is represented by the density matrix. Such a formalism
can even be applied to mixed states, which cannot be represented by a wave function.
Moreover, such a formulation in terms of partial traces allows one to construct more general Bohm-type
theories \cite{nik_solip,nik_statist} by tracing out over all degrees of freedom to which one does not assign
ontological trajectories. The theory presented in this paper can be viewed as a special case of such a more general framework developed in \cite{nik_solip,nik_statist}.

\end{document}